\documentclass{article}[12pt]
\usepackage{amssymb}
\usepackage{graphicx}

\catcode`\@=11
\def\@citex[#1]#2{%
\if@filesw \immediate \write \@auxout {\string \citation {#2}}\fi
\@tempcntb\m@ne \let\@h@ld\relax \def\@citea{}%
\@cite{%
  \@for \@citeb:=#2\do {%
    \@ifundefined {b@\@citeb}%
      {\@h@ld\@citea\@tempcntb\m@ne{\bf ?}%
      \@warning {Citation `\@citeb ' on page \thepage \space undefined}}%
      {\@tempcnta\@tempcntb \advance\@tempcnta\@ne%
      \@tempcntb\number\csname b@\@citeb \endcsname \relax%
      \ifnum\@tempcnta=\@tempcntb 
        \ifx\@h@ld\relax%
          \edef \@h@ld{\@citea\csname b@\@citeb\endcsname}%
        \else%
          \edef\@h@ld{\ifmmode{-}\else--\fi\csname b@\@citeb\endcsname}%
        \fi%
      \else
        \@h@ld\@citea\csname b@\@citeb \endcsname%
        \let\@h@ld\relax%
      \fi}%
    \def\@citea{,\penalty\@highpenalty\,}%
  }\@h@ld
}{#1}}

\def\@citeb#1#2{{[#1]\if@tempswa , #2\fi}}
\def\@citeu#1#2{{$^{#1}$\if@tempswa , #2\fi }}
\def\@citep#1#2{{#1\if@tempswa , #2\fi}}

\def\bcites{         
        \catcode`\@=11
        \let\@cite=\@citeb
        \catcode`\@=12
}

\def\upcites{         
        \catcode`\@=11
        \let\@cite=\@citeu
        \catcode`\@=12
}

\def\plaincites{      
        \catcode`\@=11
        \let\@cite=\@citep
        \catcode`\@=12
}

\newcount\hour
\newcount\minute
\newtoks\amorpm
\hour=\time\divide\hour by 60
\minute=\time{\multiply\hour by 60 \global\advance\minute by-\hour}
\edef\standardtime{{\ifnum\hour<12 \global\amorpm={am}%
        \else\global\amorpm={pm}\advance\hour by-12 \fi
        \ifnum\hour=0 \hour=12 \fi
        \number\hour:\ifnum\minute<10 0\fi\number\minute\the\amorpm}}
\edef\militarytime{\number\hour:\ifnum\minute<10 0\fi\number\minute}

\def\draftlabel#1{{\@bsphack\if@filesw {\let\thepage\relax
   \xdef\@gtempa{\write\@auxout{\string
      \newlabel{#1}{{\@currentlabel}{\thepage}}}}}\@gtempa
   \if@nobreak \ifvmode\nobreak\fi\fi\fi\@esphack}
        \gdef\@eqnlabel{#1}}
\def\@eqnlabel{}
\def\@vacuum{}
\def\marginnote#1{}
\def\draftmarginnote#1{\marginpar{\raggedright\scriptsize\tt#1}}
\def\draft{
        \pagestyle{plain}
        \overfullrule=2pt
        \oddsidemargin -.5truein
        \def\@oddhead{\sl \phantom{\today\quad\militarytime} \hfil
        \smash{\Large\sl DRAFT} \hfil \today\quad\militarytime}
        \let\@evenhead\@oddhead
        \let\label=\draftlabel
        \let\marginnote=\draftmarginnote
        \def\ps@empty{\let\@mkboth\@gobbletwo
        \def\@oddfoot{\hfil \smash{\Large\sl DRAFT} \hfil}
        \let\@evenfoot\@oddhead}
        \def\@eqnnum{(\theequation)\rlap{\kern\marginparsep\tt\@eqnlabel}%
        \global\let\@eqnlabel\@vacuum}  }

\def\eqalign#1{\null\,\vcenter{\openup\jot\m@th
  \ialign{\strut\hfil$\displaystyle{##}$&$\displaystyle{{}##}$\hfil
      \crcr#1\crcr}}\,}
\def\eqalignno#1{\displ@y \tabskip\centering
  \halign to\displaywidth{\hfil$\@lign\displaystyle{##}$\tabskip\z@skip
    &$\@lign\displaystyle{{}##}$\hfil\tabskip\centering
    &\llap{$\@lign##$}\tabskip\z@skip\crcr
    #1\crcr}}

\def\section{\@startsection {section}{1}{\z@}{3.ex plus 1ex minus
 .2ex}{2.ex plus .2ex}{\large\bf}}
\def\subsection{\@startsection{subsection}{2}{\z@}{2.75ex plus 1ex minus
 .2ex}{1.5ex plus .2ex}{\bf}}

\def\abstract{\if@twocolumn
\section*{Abstract}
\else 
\begin{center}
{\bf Abstract\vspace{-.5em}\vspace{0pt}}
\end{center}
\quotation
\fi}

\catcode`\@=12

\catcode`\@=11
\def\theequation{\arabic{equation}}
\catcode`\@=12

\bcites

\catcode`\@=11
\def\theequation{\thesection.\arabic{equation}}
\@addtoreset{equation}{section}
\@addtoreset{footnote}{section}
\@addtoreset{footnote}{subsection}
\catcode`\@=12

\newcommand{\beq}{\begin{equation}}
\newcommand{\beqa}{\begin{eqnarray}}
\newcommand{\bega}{\begin{array}}
\newcommand{\ea}{\end{array}}

\newcommand{\eeq}{\end{equation}}
\newcommand{\eeqa}{\end{eqnarray}}
\newcommand{\p}{\partial}

\newcommand{\Or}{{\cal O}}

\newcommand{\IR}{{\mathbb{R}}}
\newcommand{\IZ}{{\mathbb{Z}}}

\newcommand{\CV}{{\cal V}}
\newcommand{\CVp}{{{\cal V}^\prime}}
\newcommand{\bt}{{\tilde b}}
\newcommand{\NN}{{\cal N}}
\newcommand{\ct}{{\tilde c}}

\newcommand{\ra}{{\rightarrow}}
\newcommand{\ie}{{{\it i.e.}~}}

\begin{document} \large

\begin{titlepage}

\begin{center}
\hfill  hep-th/0507002\\

\vskip 2.5 cm
{\LARGE \bf The Ground Ring of $N=2$ Minimal String Theory}
\vskip 1 cm
{Anatoly Konechny, Andrei Parnachev and David A. Sahakyan}\\
\vskip 0.5cm
{\sl Department of Physics, Rutgers University,\\
 126 Frelinghuysen rd., Piscataway NJ 08854, USA\\ }

\end{center}

\vskip 0.5 cm

\begin{abstract}
We study the $\NN=2$ string theory or the $\NN=4$ topological string
on the deformed CHS background.
That is, we consider the $\NN=2$ minimal model coupled to the $\NN=2$ Liouville theory.
This model describes holographically the topological sector of Little String Theory.
We use degenerate vectors of the respective $\NN=2$ Verma modules
to find the set of BRST cohomologies at ghost number zero--the ground ring,
and exhibit its structure.
Physical operators at ghost number one constitute a module of the ground ring, so
the latter can be used to constrain the S-matrix of the theory.
We also comment on the inequivalence of BRST cohomologies of the $\NN=2$ string theory
in different pictures.

\end{abstract}

\end{titlepage}

\section{Introduction}
The ground ring proved to be a powerful tool in constraining the dynamics of
non-critical bosonic and $\NN=1$ fermionic string theories. In \cite{Lian:1991gk}
Lian and Zuckerman constructed BRST cohomologies at various ghost numbers.
A particularly interesting subset of
these BRST invariant operators is a set of operators of ghost number zero.
Under the operator product these operators form a ring, which is called the ground ring.
Its importance  was first realized in \cite{Witten:1991zd}, where
the relations to the symmetries of the theory were revealed.
More recently the relations between the ground ring and the D-branes of the
theory were understood \cite{Seiberg:2003nm}. In particular it was shown that
boundary states based on the ZZ branes \cite{Zamolodchikov:2001ah} turn out to be the
eigenvectors of the ground ring generators. 
Moreover, the corresponding eigenstates
label the singular points in the moduli space of boundary couplings.

An interesting set of string theories is obtained 
by gauging $\NN=2$ superconformal symmetry on the worldsheet.
The resulting model is called the $\NN=2$ string 
\cite{Ooguri:1991fp,Ooguri:1991ie}.
The ground ring of the critical $\NN=2$
string was studied in \cite{Lechtenfeld:1999ik}.
In this paper we study the ground ring of the $\NN=2$ minimal string theory. 
This model is obtained by coupling  the SU(2)/U(1) superparafermions to the $\NN=2$
Liouville theory, or equivalently to the $\NN=2$ $SL(2)/U(1)$ supersymmetric coset \cite{Hori:2001ax}.
We will use these equivalent descriptions interchangeably in the paper.
One of the main motivations for our work is better understanding of the
topological sector of Little String Theory (LST).
LST appears in various decoupling limits of string theories which contain  NS5
branes or singularities. 
An interesting property of LST is that while being a non-local 
theory it does not contain gravity.
This theory was extensively studied (for recent review and further references
see \cite{Aharony:1999ks,Kutasov:2001uf})
using the holographically dual description \cite{Aharony:1998ub},
and many interesting features have been unveiled 
\cite{Berkooz:2000mz,Harmark:2000hw,Kutasov:2000jp,Aharony:2004xn,Parnachev:2005hh}.
The simplest of LSTs are $5+1$ dimensional theories with sixteen supercharges.
They arise from the decoupling limit of $k$ type $IIA$ or type $IIB$
$NS5$-branes in flat space. The holographic description of these theories is given
by closed strings in the near horizon geometry of NS5 branes--the CHS background
\cite{Callan:1991at}.  Unfortunately, string theory in this background is strongly
coupled due to the presence of the linear dilaton. One way to avoid this problem is to consider
the theory at a non-singular point in the moduli space. The simplest such
configuration corresponds to NS5 branes distributed on a circle. In this case
the CHS background gets deformed into \cite{Giveon:1999px}
\beq\label{CHSone}
\IR^{5,1}\times \left({SU(2)\over U(1)}\times {SL(2)\over U(1)}\right)/\IZ_k,
\eeq
where the $\IZ_k$ orbifolding ensures the R-charge integrality, \ie
imposes the GSO projection.

String theory in the CHS background (and its deformation) enjoys $\NN=4$
worldsheet supersymmetry, hence one can define $\NN=4$ topological string
theory in this background, which is equivalent to the $\NN=2$ string  \cite{Berkovits:1994vy}.
This topological string captures the sector of string theory in the 
background (\ref{CHSone}), which is protected by supersymmetry.
This is similar to the description of the BPS sector of type II string theory compactified
on Calabi-Yau threefold by the $\NN=2$ topological string.
One expects that the $\NN=2$ string should holographically describe
the topological version of Little String Theory.
In \cite{Aharony:2003vk} the validity of this proposal was checked by an explicit
calculation of $F^4$ coupling, where $F$ is the Abelian gauge field in the low-energy theory, using
Heterotic/Type II duality. 

One can use the ground ring to effectively compute amplitudes of physical operators
in the $\NN=2$ string theory. Indeed, BRST invariant operators at ghost number one form a module of the ground
ring and one can use this fact to express correlation functions of these operators in terms of
the structure coefficients of the
ground ring. Let us label the physical operators of the theory $\cal T$ by multi-index $[I]$.
Then ${\cal T}_{[I]}=\Or_{[I]}{\cal T}_{[0]}$, where $\Or_{[I]}$ is a ground ring element and we can write
\beq
\langle{\cal T}_{[I]}{\cal T}_{[J]}{\cal T}_{[K]}\rangle=N_{IJK}\langle{\cal T}_{[0]}{\cal T}_{[0]}{\cal T}_{[0]}\rangle.
\eeq
In this expression $N_{IJK}$ are the structure constants of the ground ring.
Unfortunately for the bosonic and $\NN=1$ string the existence of contact terms \cite{Green:1987qu}
makes simple generalization of this result to four and higher point amplitudes hard.
This is because one has to integrate over the worldsheet position of at least one of the operators.
Hence, the explicit knowledge of the contact terms is required.
It is plausible that bigger symmetry of the $\NN=2$ string theory will allow one
to understand the structure of contact terms, and possibly
to generalize this method to four and higher-point amplitudes.

This paper is organized as follows.
In Section 2 we define the $\NN=2$ Minimal String Theory and provide necessary details.

In Section 3  the general construction of ground ring elements is presented.
It is similar to that employed in bosonic and $\NN=1$ fermionic minimal string theories.
We establish that there is a one to one correspondence between the f-series degenerate vectors of Verma modules
of $SU(2)/U(1)$ and $SL(2)/U(1)$ supercosets and ground ring elements in the
$(-1,-1)$ picture of the $\NN=2$ string.
Ground ring elements are labeled by the level at which the null vector appears in the corresponding superconformal family
and a certain quantum number $m$ which is related to the R-charge of the $SU(2)/(1)$ coset.

In Section 4 we explicitly construct ground ring elements based on the level one degenerate
operators in the $SU(2)/U(1)$ and $SL(2)/U(1)$ supercosets and show that they generate a $\IZ_k$ subring. 
This subring acts on the elements of ground ring by shifting $m$.

In Section 5 we consider degenerate vectors which give rise to null vectors at level two.
We find the explicit form of null vectors and use them to construct the level 
two ground ring elements. We argue that they are the generators of the ground ring.

In Section 6 ground ring operators based on g-series degenerate vectors are discussed.
We show that the operators responsible for the spectral flow symmetry of the theory
belong to this category along with the unity operator. These operators have total picture 0.
We argue that these are the only independent ground ring elements at this picture based on g-series.

We conclude with the discussion of our results and directions for future research.

Appendix A provides  necessary information about the $\NN=2$ superconformal ghosts.
In Appendix B we present some useful facts about $SU(2)/U(1)$ and $SL(2)/U(1)$ supercosets and
derive the fusion rules of the degenerate operators at level one. 
In Apendix C we give a detailed description of the fusion of the level one ground ring operators.
In Appendix D we give an example of a kernel and a co-kernel of a picture-raising operator,
hence illustrating the inequivalence of different pictures in the $\NN=2$ string theory.

\section{ $\NN=2$ minimal strings}
The worldsheet matter content of the $\NN=2$ minimal string theory  is given by
the product of two $\NN=2$ superconformal cosets $SL(2)/U(1)\times SU(2)/U(1)$.
The level of both cosets is a positive integer $k$. 
The corresponding (coset) central charges are
\begin{equation} \label{cc}
c_{SU} = 3\left(1 - \frac{2}{k}\right)\, , \qquad
c_{SL} = 3\left(1+  \frac{2}{k}\right) \, .
\end{equation}

The $\NN=2$ superconformal algebra reads
\begin{eqnarray}
&& [L_{m},G^{\pm}_{r}]=\left(\frac{m}{2}-r\right)G^{\pm}_{m+r} \, ,\nonumber \\
&& {[}L_{m}, J_{m}]=-nJ_{m+n} \, , \nonumber\\
&& \{G^{+}_{r},G^{-}_{s}\}=2L_{r+s} + (r-s)J_{r+s} + \frac{c}{3}\left(r^{2}-
 \frac{1}{4}\right)\delta_{r, -s} \, , \\
&& \{G^{+}_{r}, G^{-}_{s}\} = \{G^{-}_{r}, G^{-}_{s}\} = 0\, , \nonumber\\
&& {[}J_{n}, G_{r}^{\pm}]=\pm G^{\pm}_{r+n} \, , \nonumber\\
&& {[}J_{m}, J_{n}]=\frac{c}{3}m\delta_{m, -n}\,,\nonumber
\end{eqnarray}
where $L_{n}$ are the Virasoro algebra generators, $G^{\pm}_{r}$,
are the modes of the supercurrents ($r\in {\mathbb Z}+1/2$ in the Neveu-Schwarz and  $r\in {\mathbb Z}$ in the Ramond
sectors) and $J_{n}$ are modes of the $U(1)$ R-current.

Upon gauging the world-sheet $\NN=2$ superconformal algebra one introduces
the usual fermionic $(b,c)$ conformal ghost system, a pair of bosonic
$(\beta^{\pm}, \gamma^{\mp})$ systems and an additional fermionic $(\tilde b, \tilde c)$
system with weights $(0,1)$    that arises when gauging the R-current.
Our conventions regarding the ghost systems are collected in appendix A.
The  BRST current takes the form
\begin{eqnarray}
j_{BRST} = cT +  \eta_{-}e^{\phi_{-}}G^{+} +
\eta_{+}e^{\phi_{+}}G^{-}
 + \tilde cJ^{m} + \nonumber \\
\frac{1}{2}[cT^{gh} + \eta_{-}e^{\phi_{-}}G^{+}_{gh}
+ \eta_{+}e^{\phi_{+}}G^{-}_{gh} +  \tilde cJ^{gh}]\,,
\end{eqnarray}
where $T^{gh}$, $G^{\pm}_{gh}$, $J^{gh}$ are the ghost $\NN=2$ currents whose
explicit expressions are given in (\ref{ghost_currents}), $\eta^{\pm}$ and
$\phi_{\pm}$ are bosonized superghosts (see appendix A).
The total central charge of the $\NN=2$ ghost system equals -6; this  matches 
the total central charge of the matter theory (\ref{cc}).

The chiral BRST operator
\beq
Q_{BRST} = \frac{1}{2\pi i} \oint dz\, j_{BRST}\,,
\eeq
commutes with the ghost number current
\beq
\label{ghnum}
  j_{gh}=-bc-\tilde b \tilde c +\eta^+\xi^-+\eta^-\xi^+\,,
\eeq
and the two picture number currents
\beq
\label{picnum}
  j_{\pi+}=-\eta^+\xi^--\p\phi_+;\qquad j_{\pi-}=-\eta^-\xi^+-\p\phi_- \, .
\eeq
The corresponding cohomology groups are thus labeled by the ghost number and  the picture
numbers $(\Pi_{+}, \Pi_{-})$.

 One can  define two picture raising operators
 \beq\label{pco}
PCO^{\pm}=\{Q,\xi^\pm\}=
c\p\xi^\pm+e^{\phi^{\mp}}(G^{\pm}-2\eta^\pm e^{\phi^\pm}b\pm 2\p(\eta^\pm e^{\phi^\pm})
\tilde b\pm\eta^\pm e^{\phi^\pm}\p\tilde b ) \, .
\eeq
It is known \cite{Lechtenfeld:1999ik} that unlike in the $\NN=1$ superstrings, the
picture raising operators (\ref{pco}) are not isomorphisms of the absolute cohomology groups
at different pictures (see appendix D for explicit examples of the $PCO^{\pm}$
kernel and cokernel elements). 
This complicates the analysis of the cohomologies.

The $SU(2)/U(1)$ superparafermion primaries ${\cal V}_{j,m}$
are labeled by $j\in \frac{1}{2}{\mathbb Z}$, $0\le j\le \frac{k-2}{2}$
and a number $m\in \{  -j, -j +1, \dots , j-1,j\}$.
In the NS sector these primaries have conformal dimensions
\beq\label{dimsu}
  \Delta[{\cal V}_{j;m}]={j(j+1)\over k} - {m^2\over k}\,
\eeq
and the $R$-charge
\beq
\label{uonersu}
  q({\cal V}_{j,m})=-{2 m\over k} \, .
\eeq

The $SL(2)/U(1)$ $\NN=2$ primaries are denoted by ${\cal V}_{h,m}'$. They are labeled by
two real numbers $h$ and $m$. The latter number takes the values $m\in h + {\mathbb Z}_{+}$ for
a lowest weight representation, $m\in -h + {\mathbb Z}_{-}$ for a highest weight one
and $m\in -h, -h +1, \dots ,h-1,  h$ for a finite one.
The conformal dimension reads
\beq\label{dimsl}
  \Delta[{\cal V}_{h,m}']=-{h(h-1)\over k} + {m^2\over k}\,
\eeq
while the R-charge is given by
\beq
\label{uonersl}
  q({\cal V}_{h,m}')={2 m\over k} \, .
\eeq


\section{Construction of ground ring elements}
As explained above, the ground ring consists of physical states at ghost number 0.
In this section we describe the construction of nontrivial elements
of the ground ring following \cite{Imbimbo:1991ia}, where
(p,q) minimal models coupled to Liouville
theory were studied.
The minimal models contain a large set of null vectors in the Verma
module; each of them can be dressed with a null vector in the Liouville
theory to produce a physical state at ghost number 1.
It has been argued \cite{Imbimbo:1991ia} that these null vectors should be
set to zero in the physical theory.
Hence, the ghost number 1 physical state described above should be set
to zero.
Its BRST pre-image is a non-trivial physical state at ghost number 0 \cite{Imbimbo:1991ia},
otherwise known as a ground ring element.

Let us review the structure of the null vectors in $SU(2)/U(1)$ and
$SL(2)/U(1)$, following \cite{Boucher:1986bh} (see also \cite{Kiritsis:1986rv})
We first consider $SU(2)/U(1)$ at level $k$ ($SL(2)/U(1)$ expressions are
obtained by the substitution $k\ra -k$)
Here we only discuss the NS sector.
The R sector can be obtained from it by the spectral flow.
The ground ring elements which realize this operation
are described below.

The corresponding Kac determinant has two series of zeroes specified by
two functions $f_{r,s}$ and $g_{l}$ \cite{Boucher:1986bh}:
\beq
\label{fseries}
  f_{r,s}=2 (\ct-1) \Delta -q^2-{1\over4}(\ct-1)^2+{1\over4}[(\ct-1)r+2 s]^2\,,
\eeq
and
\beq
\label{gseries}
g_{l}=2\Delta-2 l q+(\ct-1)(l^2-{1\over4})\,,
\eeq
where
\beq
\ct-1=\pm{2\over k}\,,
\eeq
where the $-(+)$ sign is for $SU(2)/U(1)$ ($SL(2)/U(1)$.
The null states are at the level $rs$ for the $f$-series and at the level $|l|$ for
the $g$-series states.
The R-charge of the null states differs from that of the
corresponding primaries by zero in the case of $f$-series, and by
$\pm1={\rm sign}(l)$ for the $g$-series.

We want to construct $(-1,-1)$ picture physical states at ghost
number 1.
A natural ansatz would be
\beq
\label{anza}
   c e^{-\phi_+} e^{-\phi_-} \{-rs\} {\cal V}_{j,m_{su}}  {\cal V}'_{h,m_{sl}}\,,
\eeq
where $\{-rs\}$ stands for a raising operator at level $rs$ acting
on the $SU(2)$ superparafermion primary ${\cal V}_{j,m_{su}}$ and the $SL(2)$ superparafermion
primary ${\cal V}'_{h,m_{sl}}$.
The condition of vanishing R-charge implies $m_{su}=m_{sl}$, so
that $q_{su}=-q_{sl}$.
According to (\ref{fseries}) and (\ref{gseries}), the dimension of the matter part in
(\ref{anza}) is zero for f-series and $|l|$ for the $g$-series.
Hence, it is the $f$-series which is well suited for the construction
of physical states of the type (\ref{anza}).
Below we will identify the quantum numbers which correspond to
the zeroes of (\ref{fseries}).
We will discuss the $g$-series later in the paper.

The quantum numbers of degenerate states whose conformal dimension
and R-charge satisfy (\ref{fseries}) can be obtained with the help
of (\ref{dimsl}), (\ref{uonersl}) for $SL(2)/U(1)$ theory and
(\ref{dimsu}), (\ref{uonersu}) for $SU(2)/U(1)$.
The result for the latter is $j_{r,s}={k s-1-r\over2}$.
Since $j$ is restricted to lie between $0$ and $(k-2)/2$,
we must set $s=1$.
Hence, degenerate states in the  supersymmetric $SU(2)/U(1)$
which have a null descendant at level $s$ are parameterized by
\beq
\label{jmrs}
  j_r={k -1-r\over2},\quad r=1,\ldots, k-1;\qquad m=-j_r,-j_r+1,\ldots, j_r\,.
\eeq
Corresponding degenerate states in the supersymmetric $SL(2)/U(1)$ are
labeled by
\beq
\label{hmrs}
  h_r=-{k-1+r\over 2},\quad r=1,\ldots, k-1;\qquad m=h_r,h_r+1,\ldots, -h_r\,.
\eeq

We now describe the construction of a ground ring operator
based on the $f$-series degenerate operators $\CV_{j_r m}$,
$\CVp_{h_r m}$, with $j_r$ and $h_r$ given by
(\ref{jmrs}) and (\ref{hmrs}), respectively.
The physical state operator at ghost number 1, which should
be set to zero, has the following general form
\beq
\label{gnone}
   \Or'=c e^{-\phi_+} e^{-\phi_-} \left(\lambda \{-rs,su\}+\lambda'\{-rs,sl\}\right)
                   \CV_{j_r m} \CVp_{h_r m}\,,
\eeq
where $\lambda$ and $\lambda'$ are arbitrary coefficients.
The corresponding state is equal to zero in the physical theory.
The ground ring element $\Or$ is found by requiring
\beq
\label{grfind}
  Q_{BRST} \Or =  \Or'\,.
\eeq
Below we consider $r=1$ and $r=2$ cases in detail.
We find that the operator $\Or$ is determined up to a rescaling
and an addition of BRST-trivial piece which corresponds to the choice
of $\lambda$ and $\lambda'$ in (\ref{gnone}).
Note that $\Or$ is a ground ring element in the $(-1,-1)$ picture.

The operators that will be useful in the following are \cite{Aharony:2003vk,Bischoff:1996bn}
\begin{eqnarray}
\label{spm}
{\cal S}^{+} \equiv e^{\tilde b c/2}e^{\phi_{+}/2-\phi_{-}/2}
       {\cal V}_{0,0}(R,-) {\cal V}_{0,0}'(R,-)\,, \\
{\cal S}^{-} \equiv e^{-\tilde b c/2}e^{-\phi_{+}/2+\phi_{-}/2}
       {\cal V}_{0,0}(R,+) {\cal V}_{0,0}'(R,+)\,.
\end{eqnarray}
These operators are in the BRST cohomology of the $\NN=2$ string theory at ghost number 0, and hence
are elements of the ground ring. They satisfy
\beq
{\cal S}^+{\cal S}^-\sim 1\,,
\eeq
and  map NS sector states into R sector states, thereby providing
an isomorphism between the two sectors of the theory.
That is why it is sufficient to consider only NS sector.

By squaring (\ref{spm}) 
we obtain new ground ring operators
\begin{eqnarray}
\label{spmsq}
({\cal S}^{+})^{2} \equiv e^{\tilde b c}e^{\phi_{+}-\phi_{-}}
                {\cal V}_{\frac{k-2}{2},\frac{k-2}{2}}
                {\cal V}_{\frac{k+2}{2},-\frac{k+2}{2}}' \, ,\\
({\cal S}^{-})^{2} \equiv e^{-\tilde b c}e^{-\phi_{+}+\phi_{-}}
             {\cal V}_{\frac{k-2}{2},-\frac{k-2}{2}}
             {\cal V}_{\frac{k+2}{2},\frac{k+2}{2}}'   \, .
\end{eqnarray}
in the (1,-1) and (-1,1) pictures, respectively.
It is not hard to see that these operators have zero dimension
and R-charge.
Their existence implies that BRST cohomologies at
picture $(\Pi_+,\Pi_-)$ and $(\Pi_++n,\Pi_--n),\, n\in\IZ$
are isomorphic.
There exist two picture raising operators (\ref{pco}) but
they do not have an inverse in the $\NN=2$ string\footnote{An example of a kernel and a co-kernel of the
picture raising operators is discussed in the Apendix D.}.

\section{Ground ring operators at level one}

A null vector belonging to the superconformal family generated by
$|\Delta, q\rangle$ at the first level (f-series) is given by
\beq\label{deg}
|\chi\rangle=[(q-1)L_{-1}-(2\Delta+1)J_{-1}+G^{+}_{-1/2}G^{-}_{-1/2}]|\Delta, q\rangle\,,
\eeq
where $\Delta$ and $q$ satisfy
\beq\label{dims}
2(\tilde c -1)\Delta=q^2-\tilde c\,.
\eeq
In this expression  $\tilde c=c/3$ is the (rescaled) central charge
of $SL(2)/U(1)$ (upper sign) or $SU(2)/U(1)$ (lower sign) cosets
\beq
\tilde c=1\pm{2\over k} \, .
\eeq

We would like to find a state such that the action of the BRST charge on it will yield
a null vector of the form (\ref{deg}). The local operator corresponding to
such a state will belong to  the ground ring.
More precisely, we would like to find an operator $\Or$ such that
\beq
\bega{ll}
\{Q_{BRST},\Or\}=\\
\lambda c[(q^{(su)}-1)L_{-1}^{(su)}-(2\Delta^{su}+1)J_{-1}^{(su)}+{G^+_{-1/2}}^{(su)}
{G^{-}_{-1/2}}^{(su)}]\CV\CVp
e^{-\phi_+}e^{-\phi_-}+\\
\quad\quad\lambda^\prime c[(q^{(sl)}-1)L_{-1}^{(sl)}-
(2\Delta^{sl}+1)J_{-1}^{(sl)}+{G^{+}_{-1/2}}^{(sl)}{G^{-}_{-1/2}}^{(sl)}]\CV\CVp
e^{-\phi_+}e^{-\phi_-}.
\ea
\eeq
where $\CV$ and $\CVp$ are the degenerate operators at level one in $SU(2)/U(1)$ and
$SL(2)/U(1)$ supercosets
respectively and $\lambda$, $\lambda^\prime$ are proportionality coefficients.
One can further simplify the expressions above, by noting that
\beq\label{conds}
\bega{ll}
\Delta^{(sl)}=-1-\Delta^{(su)}\equiv -1-\Delta\,,\\
q^{(sl)}=-q^{(su)}\equiv -q\,,
\ea
\eeq
that follows from the physical state condition on the operator $\{Q,\Or\}$.
 These equations are consistent with the condition that degenerate operators in $SU(2)/U(1)$ and $SL(2)/U(1)$
cosets lie in the f-series at level one. We can solve the conditions (\ref{dims}), (\ref{conds})
explicitly to obtain
\begin{equation}
\CV\CVp =\CV_{\frac{k-2}{2}, m}\CVp_{-\frac{k}{2}, m} \equiv \CV_{m}^{(1)}\CVp_{m}^{(1)}
\end{equation}
with
\begin{equation}
-\frac{k-2}{2}\le m \le \frac{k-2}{2} \, .
\end{equation}

To find the operator $\Or$ we start by considering the following ansatz
\beq\label{generalO}
\bega{ll}
\Or=e^{\alpha\tilde b c} \CV^{(1)}_{m}\CVp^{(1)}_{m} e^{-\phi_+}e^{-\phi_-}+\\
\quad\quad\p\xi^-c(A^{(su)}_{m}{G^+_{-1/2}}^{(su)}+A^{(sl)}_{m}{G^+_{-1/2}}^{(sl)})
\CV^{(1)}_{m}\CVp^{(1)}_{m} e^{-2\phi_+}e^{-\phi_-}+\\
\quad\quad c\p\xi^+(B^{(su)}_{m}{G^-_{-1/2}}^{(su)}+B^{(sl)}_{m}{G^-_{-1/2}}^{(sl)})
\CV^{(1)}_{m}\CVp^{(1)}_{m} e^{-\phi_+}e^{-2\phi_-}.
\ea
\eeq
Acting by $Q_{BRST}$ we find the following independent equations on the coefficients
\beq
\bega{ll}
B^{(sl)}_{m}=A^{(su)}_{m}\equiv A_{m}\,,\\
A^{(sl)}_{m}=B^{(su)}_{m}\equiv B_{m}\,,\\
B_{m}-A_{m}=\lambda\,,\\
1+2A_{m}=\lambda(q_{m}-1)\,,\\
\alpha=-(2\Delta_{m}+1)\lambda\,.
\ea
\eeq
Solving these equations we find one parameter family of solutions
\beq\label{gensol}
\bega{ll}
2A_{m}=\lambda(q_{m}-1)-1\,,\\
2B_{m}=\lambda(q_{m}+1)-1\,,\\
\alpha=-\lambda(2\Delta_{m}+1)\,.
\ea
\eeq
At first sight this result seems to be puzzling, since
we expected to get a single solution for a fixed $m$. To resolve this puzzle we note that
$Q\Or_{\lambda=0}$ is identically zero. So it seems that we found two ``physical''
operators instead of one. It turns out, however, that $\Or_{\lambda=0}$ is actually BRST
exact. One can show that
\beq
\Or_{\lambda=0} = \{Q,  \frac{1}{2}c\partial \xi^{+}\partial \xi^{-}\CV^{(1)}_{m}\CVp^{(1)}_{m}
e^{-2\phi_+}e^{-2\phi_-} \} \, .
\eeq
(A fast way to see that $\Or_{\lambda=0}$ is cohomological to zero is by checking
$PCO^{+}\Or=0$.)

Hence indeed we obtain a single non-trivial ground ring operator corresponding to the given degenerate state.
One can conveniently fix the gauge by considering $\Or_\lambda-\Or_{\lambda=0}$ as a representative of the ground ring.
After an appropriate rescaling we obtain  ground ring operators
\beq\label{gaugc}
\bega{ll}
\Or^{(1)}_{m}=&\tilde b c \CV\CVp e^{-\phi_+}e^{-\phi_-}+\\
&\p\xi^-c(A_{m}'{G^+_{-1/2}}^{(su)}+B_{m}'{G^+_{-1/2}}^{(sl)})
\CV_{m}^{(1)}\CVp_{m}^{(1)} e^{-2\phi_+}e^{-\phi_-}+\\
&c\p\xi^+(B_{m}'{G^-_{-1/2}}^{(su)}+A_{m}'{G^-_{-1/2}}^{(sl)})
\CV_{m}^{(1)}\CVp_{m}^{(1)} e^{-\phi_+}e^{-2\phi_-}\,,
\ea
\eeq
where
\beq\label{ABp}
\bega{ll}
A_{m}'=-{(q_{m}-1)\over 2(2\Delta_{m}+1)}=\frac{1}{k-2m}\,,\\
B_{m}'=-{(q_{m}+1)\over 2(2\Delta_{m}+1)}=-\frac{1}{k+2m}\,.\\
\ea
\eeq
For certain applications it is convenient to consider these operators at higher pictures.
In particular we would like to apply $PCO^+$ to the operator (\ref{gaugc})
\beq\label{PCOp}
\bega{ll}
PCO^+\Or_{m}^{(1)}=
&\tilde b c G^{+}_{-1/2}\CV_{m}^{(1)}\CVp_{m}^{(1)}e^{-\phi_+} +
\p\xi^-{c\over(2\Delta+1)}e^{-2\phi_+}
{G^+_{-1/2}}^{(sl)}{G^+_{-1/2}}^{(su)}\CV_{m}^{(1)}\CVp_{m}^{(1)}+\\
\\
&2(A_{m}'{G^+_{-1/2}}^{(su)}+B_{m}'{G^+_{-1/2}}^{(sl)})\CV_{m}^{(1)}\CVp_{m}^{(1)}
 e^{-\phi_+}
+c\p\xi^+e^{-\phi_+}e^{-\phi_-}\CV_{m}^{(1)}\CVp_{m}^{(1)}\,.
\ea
\eeq
The $PCO^-$ for this operator reads
\beq\label{PCOm}
\bega{ll}
PCO^-\Or^{(1)}_{m}=&-\tilde b c G^{-}_{-1/2}\CV_{m}^{(1)}\CVp_{m}^{(1)}
e^{-\phi_-} +\p\xi^+ c{1\over(2\Delta+1)}e^{-2\phi_-}
{G^-_{-1/2}}^{(sl)}{G^-_{-1/2}}^{(su)}\CV_{m}^{(1)}\CVp_{m}^{(1)}+\\
\\
&-2(B_{m}'{G^-_{-1/2}}^{(su)}+A_{m}'{G^-_{-1/2}}^{(sl)})\CV_{m}^{(1)}\CVp_{m}^{(1)} e^{-\phi_-}
-c\p\xi^-e^{-\phi_+}e^{-\phi_-}\CV_{m}^{(1)}\CVp_{m}^{(1)}\,.
\ea
\eeq

The $PCO^+$ operator takes a particularly nice form when the $\CV_{m}^{(1)}$ is a chiral primary
that is when $m=1-k/2$. Then we have
\beq\label{pcoplus}
PCO^+\Or^{(1)}_{1-k/2}=
e^{-\phi_+} e^{-\tilde b c}G^{+}_{-1/2}\CV_{1-k/2}^{(1)}\CVp_{1-k/2}^{(1)}
+c\p\xi^+e^{-\phi_+}e^{-\phi_-}\CV_{1-k/2}^{(1)}\CVp_{1-k/2}^{(1)} \, .
\eeq

We can compute the OPE of two level one ground ring operators
$PCO^+\Or_{m_{1}}^{(1)}$ and $PCO^-  \Or_{m_{2}}^{(1)}$ given by (\ref{PCOp}), (\ref{PCOm}).
Thus the ground ring elements we are fusing are chosen in the gauge (\ref{gaugc}).
To compute the OPE's of the matter operators we  use
formulas (\ref{strube})-(\ref{structsl}) and similar
formulas for the $SU(2)/U(1)$ coset.


After a straightforward computation  we obtain that for $m_{1}+m_{2}<0$ we have the
following product of cohomologies
\beq
\label{pcoope}
PCO^+\Or_{m_{1}}^{(1)} \, PCO^- \Or_{m_{2}}^{(1)}=
K_{m_{1}, m_{2}}  \tilde \Or_{m_{3}}^{(1)}\,,
\eeq
where $\tilde \Or_{m_3}^{(1)}$  is the ground ring element given in
(\ref{generalO}) with $m_3=m_1+m_2+k/2$ and a $\lambda=-1/(q_1+q_2)$ fixed such that
  $B_{m_{3}}=0$ in  (\ref{gensol}). The numerical coefficient $K_{m_{1}, m_{2}}$ in
 (\ref{pcoope}) is
 \begin{equation}\label{K}
K_{m_{1}, m_{2}}=-C_{m_{1},m_{2}} {(q_1+q_2)^2 [k(q_1+q_2-2)-2][k(q_1+q_2-2)+2]\over
           k^2(q_1-1)(q_2-1)}
 \end{equation}
where $C_{m_{1},m_{2}}$  is defined as an OPE coefficient
\beq
\CV_{m_{1}}^{(1)}\CVp^{(1)}_{m_{1}}(z_1)\CV_{m_{2}}^{(1)}
\CVp_{m_{2}}^{(1)}(z_2)\sim {C_{m_{1},m_{2}}\over z_{12}^{\Delta_1+\Delta_2-\Delta_3}}
                G^{+,su}_{-1/2}\CV_{m_{3}}^{(1)} G^{-,sl}_{-1/2}\CVp_{m_3}^{(1)} \, .
\eeq
For $m_{1}+m_{2}>0$ one obtains a result similar to (\ref{pcoope}) with
an operator $\tilde {\cal O}_{m_{3}}^{(1)}$, $m_{3}=m_{1}+m_{2}-k/2$
appearing on the RHS being in a different gauge
and with a different proportionality coefficient
\begin{equation}\label{Kplus}
K^\prime_{m_{1}, m_{2}}=C_{m_{1},m_{2}}^\prime {(q_1+q_2)^2 [k(q_1+q_2-2)-2][k(q_1+q_2-2)+2]\over
           k^2(q_1+1)(q_2+1)}~,
 \end{equation}
where $C_{m_{1},m_{2}}^\prime$  is defined as follows
\beq
\CV_{m_{1}}^{(1)}\CVp^{(1)}_{m_{1}}(z_1)\CV_{m_{2}}^{(1)}
\CVp_{m_{2}}^{(1)}(z_2)\sim {C_{m_{1},m_{2}}^\prime\over z_{12}^{\Delta_1+\Delta_2-\Delta_3}}
                G^{-,su}_{-1/2}\CV_{m_{3}}^{(1)} G^{+,sl}_{-1/2}\CVp_{m_3}^{(1)} \, .
\eeq
In the case $m_{1}+m_{2}=0$ the product of the corresponding cohomologies yields the $\Or_{-{k\over 2}}$
\beq
{\cal O}^{(1)}_{-\frac{k}{2}}={\cal O}^{(1)}_{\frac{k}{2}}=c\partial\xi^{+} \CVp_{-\frac{k}{2}}^{(1)}e^{-2\phi_{-}}e^{-\phi_{+}}  +
\partial\xi^{-} c\CVp_{\frac{k}{2}}^{(1)} e^{-2\phi_{+}}e^{-\phi_{-}}\, .
\eeq
One further finds that the operator ${\cal O}^{(1)}_{-\frac{k}{2}}$ acts (up to an insignificant
 numerical constant) as the identity operator within the ground ring.
The technical details related to the fusion of operators ${\cal O}^{(1)}_{m}$ appear in
appendix C.

It is not hard to see from the above OPEs that the operators ${\cal O}_{m}^{(1)}$
generate a  cohomology subring isomorphic to $\IZ_k$
with a generator
$x={\cal O}_{1-k/2}^{(1)}$.\footnote{Note that to compute powers
of $x$ one only needs to use the product rule (\ref{pcoope}).}

\section{Ground ring operators at level two}
In this section we construct the ground ring operators at level two.
As we show below these operators play an important role in constructing the complete ground ring.
We start by computing the null vector belonging to superconformal family generated by
$|\Delta,q\rangle$, where $\Delta$ and $q$ satisfy  relation (\ref{fseries}) with $r=2$ and $s=1$
\beq
\Delta={q^2\over 2(\tilde c-1)}+{1\over 8}(\tilde c -1)-{\tilde c^2\over 2(\tilde c-1)}.
\eeq
From the Kac determinant we know that this family has a null vector at level two with the R-charge equal
to the original one. Hence we can write down the following ansatz for the null vector
\beq
\bega{ll}
 \hat f^{(2)}|\Delta,q\rangle= &(A L_{-2}+B J_{-2}+C L_{-1}^2+D J_{-1}^2+E L_{-1}J_{-1}+
F G^{+}_{-1/2}G^{-}_{-3/2}+\\
 &g G^{+}_{-3/2}G^{-}_{-1/2}+
H L_{-1}G^{+}_{-1/2}G^{-}_{-1/2}+ I J_{-1} G^{+}_{-1/2}G^{-}_{-1/2})|\Delta,q\rangle~.
\ea
\eeq
It is a straightforward exercise to compute the coefficients in this expression.
We find (up to an overall normalization)
\begin{eqnarray}\label{vec2}
A&=&1\nonumber\\
B&=&{4+3q-\ct(2+q)\over (-1+\ct)^2}\nonumber\\
C&=&{2\over -1+\ct}\nonumber\\
D&=&-{(-3+\ct-2q)(1+\ct+2q)\over 2(-1+\ct)^3}\nonumber\\
E&=&-{4(1+q)\over(-1+\ct)^2}\\
F&=&{2(-3+\ct)\ct-4q\over (-1+\ct)^2(-3+\ct+2q)}\nonumber\\
g&=&-{2(1+\ct+2q)((-3+\ct)\ct+2q)\over(-1+\ct)^2(1+\ct-2q)(-3+\ct+2q)}\nonumber\\
H&=&-{16q\over (-1+\ct)(1+\ct-2q)(-3+\ct+2q)}\nonumber\\
I&=&{16(1+\ct+2\Delta)\over (-1+\ct)(1+\ct-2q)(-3+\ct+2q)\nonumber }~.\\
\end{eqnarray}
Following the general procedure outlined in section 3, we would like to find an operator
such that the action of BRST operator yields a null physical operator at ghost number one
\beq\label{lev2eq}
\{Q,\Or^{(2)}_m\}=ce^{-\phi_+}e^{-\phi_-}(\lambda \hat f^{(2)(su)}+\lambda' \hat f^{(2)(sl)})\CV^{(2)}_m\CVp^{(2)}_m~.
\eeq
In this expression $\CV^{(2)}_m$ and $\CVp^{(2)}_m$  are $SU(2)/U(1)$ and $SL(2)/U(1)$ degenerate operators at level 2
\beq
\CV^{(2)}_m\CVp^{(2)}_m=\CV_{{k-3\over 2},m}\CVp_{-{k+1\over 2},m}\equiv \CV\CVp~.
\eeq
Now we are ready to write down an ansatz for the ground ring operator
\beq
\bega{ll}
\Or^{(2)}=c\p\tilde b \CV\CVp e^{-\phi_+}e^{-\phi_-}+a_0 cb \CV\CVp e^{-\phi_+}e^{-\phi_-}+\\
c\p^2\xi^+\{a_1^{(-)}G_{-1/2}^{-(su)}+b_1^{(-)} G_{-1/2}^{-(sl)}\}\CV\CVp e^{-\phi_+}e^{-2\phi_-}+\\
c\p\xi^+\p\phi_-\{a_{11}^{(-)}G_{-1/2}^{-(su)}+
                               b_{11}^{(-)} G_{-1/2}^{-(sl)}\}\CV\CVp e^{-\phi_+}e^{-2\phi_-}+\\
c\p\xi^+\{a_2^{(-)}L_{-1}^{(su)}G_{-1/2}^{-(su)}+b_2^{(-)}L_{-1}^{(sl)}G_{-1/2}^{-(sl)}+
c_2^{(-)}L_{-1}^{(su)}G_{-1/2}^{-(sl)}+d_2^{(-)}L_{-1}^{(sl)}G_{-1/2}^{-(su)}\\
a_3^{(-)}J_{-1}^{(su)}G_{-1/2}^{-(su)}+b_3^{(-)}J_{-1}^{(sl)}G_{-1/2}^{-(sl)}+
c_3^{(-)}J_{-1}^{(su)}G_{-1/2}^{-(sl)}+d_3^{(-)}J_{-1}^{(sl)}G_{-1/2}^{-(su)}+\\
a_4^{(-)}G_{-3/2}^{-(su)}+b_4^{(-)}G_{-3/2}^{-(sl)}                         +
\}\CV\CVp e^{-\phi_+}e^{-2\phi_-}+\\
\{terms\,\, obtained\,\, by\,\, replacing\,\, + \,\, superscript \,\, by\,\, - \}+\\
\{a_5 e^{\alpha_5 c\tilde b} L_{-1}^{(su)}+b_5 e^{\beta_5 c\tilde b}L_{-1}^{(sl)}+
a_6 e^{\alpha_6 c\tilde b} J_{-1}^{(su)}+b_6 e^{\beta_6 c\tilde b}J_{-1}^{(sl)}+\\
a_7 e^{\alpha_7 c\tilde b}G_{-1/2}^{+(su)}G_{-1/2}^{-(su)}+b_7e^{\beta_7 c\tilde b} G_{-1/2}^{+(sl)}G_{-1/2}^{-(sl)}+\\
c_7 e^{\gamma_7 c\tilde b} G_{-1/2}^{+(su)}G_{-1/2}^{-(sl)}+d_7 e^{\delta_7 c\tilde b}
G_{-1/2}^{+(sl)}G_{-1/2}^{-(su)}\}\CV\CVp e^{-\phi_+}e^{-\phi_-}~.\\
\ea
\eeq
Using the form of the null vector (\ref{vec2}) and the BRST current (\ref{BRST1}) in equation (\ref{lev2eq}) one can find the
ground ring element. The result of this calculation is quite cumbersome for general $m$ and we will not present it here.
Instead we note that one can generate all level two ground ring operators using the relation
\beq
PCO^+\Or^{(2)}_{-{k-3\over 2}}PCO^-\Or^{(1)}_m=\Or^{(2)}_{m+{3\over 2}}~,
\eeq
which can be derived using (\ref{opegpex}) (and corresponding expression for the $SU(2)$ part).
Hence for all applications it is is sufficient to know the form of the
$\Or^{(2)}_m$ for special value of  $m=-(k-3)/2$. In this case the $SU(2)/U(1)$ operator $\CV$ becomes chiral, 
which simplifies the result considerably.
The application of $PCO^+$  on $\Or^{(2)}$ yields then 
the following result\footnote{We use Mathematica to solve (\ref{lev2eq})
and find the coefficients in (\ref{pcolev2sol}).}
\beq\label{pcolev2sol}
\bega{ll}
PCO^+\Or=c\p\xi^+(2(1+k)L_{-1}^{(su)}+(1-k)L_{-1}^{(sl)}-2k J_{-1}^{(su)}+2k J_{-1}^{(sl)}+{1\over 2}G^{+(sl)}_{-1/2}G^{- (su)}_{-1/2}\\
+{k+1\over 2}G^{+(sl)}_{-1/2}G^{-(sl)}_{-1/2})\CV\CVp e^{-\phi^+}e^{-\phi^-}+\\
\\
{1\over 2}c\p\xi^-(G^{+(sl)}_{-3/2}G^{+(sl)}_{-1/2}+(k+1)G^{+(su)}_{-3/2}G^{+(sl)}_{-1/2})\CV\CVp e^{-2\phi^+}+\\
\\
(2G^{+(su)}_{-3/2}+2(1+k)G^{+(sl)}_{-3/2}+k L^{(sl)}_{-1} G^{+(sl)}_{-1/2}-k L^{(su)}_{-1} G^{+(sl)}_{-1/2}-\\
(1+2k)J^{(sl)}_{-1} G^{+(sl)}_{-1/2}+(k-1)J^{(su)}_{-1} G^{+(sl)}_{-1/2})\CV\CVp e^{-\phi^+}+\\
\\
c\tilde b(-2kG^{+(su)}_{-3/2}+2kG^{+(sl)}_{-3/2}+(k-1) L^{(sl)}_{-1} G^{+(sl)}_{-1/2}-
(1+2k) L^{(su)}_{-1} G^{+(sl)}_{-1/2}-\\
2kJ^{(sl)}_{-1} G^{+(sl)}_{-1/2}+2k J^{(su)}_{-1} G^{+(sl)}_{-1/2})\CV\CVp e^{-\phi^+}~.
\ea
\eeq
We expect from the fusion rules of the level two operators that 
\beq
PCO^+\Or^{(2)} PCO^-\Or^{(2)}\sim \Or^{(1)}+\Or^{(3)}~.
\eeq
Hence using this relation one can generate all the f-series elements of the ground ring. 
\section{BRST cohomologies related to the $g$-series }\label{gsec}
As was already noted in section 3,  the construction
of ground ring elements which correspond to null vectors in the $g$-series
(\ref{gseries}) is complicated by the fact that there is no simple general ansatz 
that would be similar to (\ref{anza}). Below we will analyze some particular ground ring
operators based on the $g$-series. Those examples show  that in the Verma module
those operators ${\cal O}$ satisfy an equation $Q_{BRST}{\cal O} = {\cal O}'$.
Here ${\cal O}'$ is a null operator which in general is given by a nontrivial linear combination
of {\it descendants} of  some primitive null vectors. This set of primitive
null vectors can contain vectors from  the $g$-series  as well as from the  $f$-series. This
is to be contrasted with the simple equation (\ref{grfind}).

\noindent \underline{\bf The identity operator} \\
We have in the matter sector
\beq
Q_{BRST} |0\rangle = cL_{-1}|0\rangle + \gamma^{-}G_{-1/2}^{+}|0\rangle  +
\gamma^{+}G_{-1/2}^{-}|0\rangle \, .
\eeq
The identity operator has $q=R=0$, $\Delta=0$. It belongs to the $g$-series of degenerate
operators and has two null vectors at level $1/2$: $G_{-/1/2}^{\pm}|0 \rangle$. The
vector $L_{-1}|0\rangle$ is a sum of descendants of these two null vectors. This follows
from the anticommutator $\{G^{+}_{-1/2}, G^{-}_{-1/2}\} = 2L_{-1}$.

\noindent \underline{\bf Operators ${\cal S}^{\pm}$} \\
Consider the operators ${\cal S}^{\pm}$ (\ref{spm}) which are ground ring elements in
the Ramond sector.
The operators ${\cal V}_{0, 0}(RR,\pm)$ and ${\cal V}_{0,0}'(RR, \pm)$  belong to the $g$=series.
The operators
\begin{equation}\label{primitive1}
G^{\pm}_{-1}{\cal V}_{0,0}(RR, \pm){\cal V}_{0,0}'(RR,\pm) \, , \qquad
G^{\mp}_{0}{\cal V}_{0,0}(RR, \pm){\cal V}_{0,0}'(RR,\pm) \, .
\end{equation}
are null.

When we check the BRST invariance of the operators ${\cal S}^{\pm}$ we obtain
\begin{eqnarray} \label{null1}
&&[Q_{BRST}, {\cal S^{\pm}}] = e^{\pm\tilde b c/2}
[\eta^{\pm}e^{3\phi_{\pm}/2-\phi_{\mp}/2}G^{\mp}_{-1}
 +
\eta^{\mp}e^{\phi_{+}/2+\phi_{-}/2}G^{\pm}_{0} + \nonumber \\
&&ce^{\pm(\phi_{+}/2-\phi_{-}/2)}(L_{-1} \pm \frac{1}{2}J_{-1})]
{\cal V}_{0,0}(RR, \mp){\cal V}_{0,0}'(RR,\mp)\, .
\end{eqnarray}
 While the first two operators in the right hand side of (\ref{null1}) are primitive
 null operators (\ref{primitive1}) the operators
\beq
(L_{-1} \pm \frac{1}{2}J_{-1}){\cal V}_{0,0}(RR, \mp){\cal V}_{0,0}'(RR,\mp)
\eeq
can be represented as  sums of descendants of the  null vectors (\ref{primitive1}). This follows
from the anticommutators
\beq
\{G^{+}_{0}, G^{-}_{-1}\}= 2L_{-1} + J_{-1}\, \quad
\{G^{+}_{-1}, G^{-}_{0}\}= 2L_{-1} - J_{-1} \, .
\eeq

\noindent \underline{\bf Operators $({\cal S}^{\pm})^{2}$} \\
Consider the operators $({\cal S}^{\pm})^{2}$ given in (\ref{spmsq}).
The primaries
\beq
{\cal V}_{\pm}\equiv {\cal V}_{\frac{k-2}{2},\pm \frac{k-2}{2}} \, , \quad
 {\cal V}_{\pm}'\equiv {\cal V}_{\frac{k+2}{2},\pm \frac{k+2}{2}}'
 \eeq
 belong simultaneously to the $g$-series where
 they have
a null vector at level $1/2$ and to the $f$-series where they have a null vector at level 1.
The corresponding null operators read
\begin{eqnarray}\label{nullvv}
&&G_{-1/2}^{\pm}{\cal V}_{\mp} \, , \quad G_{-1/2}^{\pm}{\cal V}_{\pm}' \, , \nonumber \\
&& [2(\frac{1}{k}-1)(L_{-1} \pm J_{-1}) + G^{\pm}_{-1/2}G_{-1/2}^{\mp}]
{\cal V}_{\pm} \, ,\nonumber \\
&& [-2(\frac{1}{k}+1)(L_{-1} \pm J_{-1}) + G^{\pm}_{-1/2}G_{-1/2}^{\mp}]{\cal V}_{\mp}' \, .
\end{eqnarray}
Note that the $f$-series vectors are up to the $1/2$-descendants of the $g$-series vectors
 just $(L_{-1} \pm J_{-1}){\cal V}_{\pm}$,
 $(L_{-1} \pm J_{-1}){\cal V}_{\mp}'$.

 We have
\begin{eqnarray}
&&Q_{BRST}({\cal S}^{\pm})^{2}= e^{\pm \tilde b c}[\eta^{\pm}e^{2\phi_{\pm}-\phi_{\mp}}
G^{\mp}_{-3/2}
{\cal V}_{\pm}{\cal V}_{\mp}' + (\partial \eta^{\pm} + \partial \phi_{\pm}\eta^{\pm})
e^{2\phi_{\pm}-\phi_{\mp}}G^{\mp}_{-1/2}{\cal V}_{\pm}{\cal V}_{\mp}' + \nonumber\\
&& c e^{\pm(\phi_{+}-\phi_{-})}(L_{-1} \pm J_{-1}){\cal V}_{\pm}{\cal V}_{\mp}']\, .
\end{eqnarray}
The second and the third terms in the RHS are null. This follows directly from the
form of the primitive null vectors (\ref{nullvv}).
 Whereas the first term can be represented
as a sum of descendants of the $g$ and $f$-series  null vectors
(\ref{nullvv}) because of the commutation
relations
\beq
[L_{-1} + J_{-1}, G^{-}_{-1/2}] = - G^{-}_{-3/2} \, , \quad
[L_{-1} - J_{-1}, G^{+}_{-1/2}] = - G^{+}_{-3/2} \, .
\eeq

Consider now a general situation.
From (\ref{gseries}) we find that  any primary ${\cal V}_{j,m}$ belongs to the $g$-series
and has two null vectors at levels $l=j + \frac{1}{2} \pm m$ while a primary
${\cal V}_{h, m'}'$ belongs to the $g$ series if $l= m' + \frac{1}{2} \pm h \in
\frac{1}{2}{\mathbb Z}$.
The corresponding $SL(2)$ representation based on ${\cal V}'_{h,m'}$  is finite if $h<0$.
The condition that the total $R$-charge of the operator
$ {\cal V}_{j, m}{\cal V}_{h, m'}$ is an integer $R$ implies that $m-m' = \frac{k}{2}R$.
The total dimension of this operator is
\beq
\Delta = \frac{(j+h)(j-h+1)}{k} + \frac{(m + m')R}{2} \,.
\eeq
Generically for $\Delta$ to be an integer or a half-integer one has to set
$h + j = 0$ or $h+j = -k/2$ (note the unitarity bounds on $h$ and $j$).
Or alternatively one can have the reflected equations with $h$ replaced by $1-h$.
In the remainder of this section we will concentrate on the case $h=-j$.
(Cohomologies based on the matter primaries with $h+j=-k/2$ can be obtained by applying
the spectral flow to the cohomologies based on the $f$-series considered in the previous
two sections.) Since $|m|\le j$, $|m'|\le h$  it follows that in $m=m'$ in this case.
 Denote such a primary
${\cal V}{\cal V}'\equiv {\cal V}_{j, m}{\cal V}_{-j, m}$. Its total $R$-charge and
dimension are zero. The prescribed picture numbers $\Pi_{\pm}=0$ and the vanishing total
$R$-charge allow for an operator of the form
\beq
{\cal O} = F(b,\tilde b, c, \tilde c) (\gamma^{+})^{N_{+}}(\gamma_{-})^{N_{-}}
(\beta^{+})^{M_{+}}(\beta^{-})^{M_{-}}{\cal V}{\cal V}' \, .
\eeq
The $R$-charge conservation requires $N_{+} - N_{-} -M_{+} + M_{-} = 0$.
We further note that the combinations $\gamma^{-}\beta^{+}$ and $\gamma^{+}\beta^{-}$
do not carry any charges except for the conformal weight $1$. Since there are no
operators in the ghost Fock space with ghost number zero and negative conformal dimension
the above two combinations can be dropped. Since $\beta^{\pm}$ have positive conformal
dimension this leaves us with the powers $(\gamma^{+}\gamma^{-})^{N}$. The latter
have ghost number $2N$ and conformal dimension $-N$. One finally notes that
there are no combinations of the remaining ghosts $b,c,\tilde b, \tilde c$ that can compensate
both of those charges. We conclude that $N=0$. The only uncharged combination of the remaining
ghosts is $c\tilde b$.
The upshot of these considerations is that the only ghost
structure  that can be
present in a nontrivial cohomology representative based on the ${\cal V}{\cal V}'$ is
$c\tilde b$.
(The above discussion concerned the primaries themselves,  by similar considerations
one can conclude that no descendant of such a family can be present in a nontrivial cohomology.)
By acting with the BRST charge $Q_{BRST}$ on an operator of the form
\beq
e^{\alpha c\tilde b}{\cal V}{\cal V}'
\eeq
we obtain
\beq
e^{\alpha c\tilde b}[\gamma^{+}G_{-1/2}^{-} + \gamma^{-}G_{-1/2}^{+}]{\cal V}{\cal V}'
+  cL_{-1}{\cal V}{\cal V}' -  \alpha c J_{-1}{\cal V}{\cal V}'\, .
\eeq
The right hand side of this expression has a zero norm only if ${\cal V}{\cal V}'$ is
the identity operator and $\alpha = 0$. Thus we see that (up to the operators obtained
by applying the spectral flow to the $f$-series based cohomologies) the identity
operator is essentially the only cohomology in picture $(0,0)$ based on the $g$-series.


\section{Conclusions and future directions}
In this paper we studied BRST cohomologies at ghost number zero in the $\NN=2$
minimal string theory. Such cohomologies are related to null vectors present in the
$SU(2)/U(1)$ and $SL(2)/U(1)$ parafermion representation spaces on which one
builds up the matter sector of the theory. We explicitly constructed the ground ring
elements corresponding to level one and level two primitive null states belonging to
the $f$-series of degenerate representations. The superparafermion fusion
rules together with our analysis of the cohomologies based on the $g$-series
imply that these elements generate all of the ground ring\footnote{There can still
be some accidental cohomologies based on the $g$-series at higher pictures, but
we do not think those can be of much practical value}.
A particularly simple subring isomorphic to the ${\mathbb Z}_{k}$ ring, where
$k$ is the level of parafermion algebras, is generated by the cohomologies based
on the degenerate operators with the null states at level one.
This construction is described in detail in section 4 of
this paper.

As mentioned in the introduction, the study of the ground ring
in the $\NN=2$ minimal string theory
presented in this paper is motivated by a number of potential
applications.  One of them is the structure of topological amplitudes on the CHS background.
To be able to apply the ground ring in this context one needs to investigate the structure of
the module over the ground ring that is carried by the physical operators with ghost number one.
Possible contact terms in the correlators also need to be understood.
We hope to address these issues and to compute some amplitudes explicitly by this method
in the future.

Another interesting direction is extending the interrelations between D-branes, the ground ring
and a spectral curve found in \cite{Seiberg:2003nm} to the $\NN=2$ minimal strings.
It was found in \cite{Seiberg:2003nm} that the geometrical object behind minimal bosonic
and minimal type 0  strings is a complex curve parameterized by a complexified
boundary Liouville constant. The singularities of this curve are in correspondence with
the Liouville theory ZZ-branes \cite{Zamolodchikov:2001ah} on the one hand and with
the ground ring and its module relations
on the other.
D-branes in $\NN=2$ super Liouville theory were studied in a number of papers
\cite{Ahn:2004qb,Hosomichi:2004ph,Troost_etal,Eguchi}.
The relations between the $\NN=2$ analogs of the ZZ-branes found in \cite{Ahn:2004qb}
and their FZZT-brane \cite{FZZT} counterparts suggest that 
in the $\NN=2$ case there is a complex surface
analogous to the spectral curve of \cite{Seiberg:2003nm}. We will report on the
details of its construction and the relations with the ground ring
elsewhere \cite{KPS2}. Elucidating this structure potentially may  be useful for
 finding a matrix model description of the $\NN=2$ two-dimensional black hole
and little strings.

\noindent {\bf \Large Acknowledgements}

We are grateful to D. Belov, D. Friedan, D. Kutasov and G. Moore for useful discussions.
This work was partially supported by DOE grant DE-FG02-96ER40949.


\appendix
\renewcommand{\theequation}{\Alph{section}.\arabic{equation}}
\setcounter{equation}{0}

\section{$\NN=2$ superconformal ghosts }

\begin{eqnarray}\label{ghost_currents}
&&T^{gh} = 2\partial c b+c\partial b + \partial \tilde c\tilde b -\frac{3}{2}
(\partial \gamma^{-}\beta^{+} + \partial\gamma^{+}\beta^{-})
-\frac{1}{2}(\gamma^{-}\partial \beta^{+} + \gamma^{+}\partial \beta^{-}) \, , \nonumber \\
&&G^{\pm}_{gh} =-2\gamma^{\pm} b \pm 2\partial \gamma^{\pm}\tilde b \pm 2\gamma^{\pm}\partial \tilde b
+ \frac{3}{2}\partial c \beta^{\pm} + c\partial \beta^{\pm} \pm \tilde c\beta^{\pm} \, ,
\nonumber \\
&&J^{gh} =\partial (c\tilde b) + \gamma^{-}\beta^{+} - \gamma^{+}\beta^{-} \, .
\end{eqnarray}
These operators satisfy $N=2$ algebra \cite{Polchinski:1998rr}.
The super ghosts  OPE reads
\begin{equation}
\beta^{\pm}(z)\gamma^{\mp}(w) \sim \frac{-1}{z-w}
\end{equation}
We bosonise superconformal ghosts as
\beq
\label{bgbos}
   \beta^\pm\cong e^{-\phi_\mp} \p\xi^\pm;\qquad  \gamma^\pm\cong\eta^\pm e^{\phi_\pm}
\eeq
which implies
\beq
   \eta^\pm(z) \xi^\mp(0)\sim {1\over z}
\eeq
In (\ref{bgbos}) $\phi$ is the canonically normalized scalar.
It is sometimes convenient to express the BRST current in terms of bosonised ghosts
\beq\label{BRST1}\bega{ll}
J_{BRST}=&cT+\eta^+e^{\phi_+}G^-+\eta^-e^{\phi_-}G^++\tilde c J+\\
&c[\p c b+\p\tilde c\tilde b-{1\over 2}(\p\phi^+)^2-{1\over 4}\p^2\phi^+-{1\over 2}(\p\phi^-)^2-{1\over 4}\p^2\phi^--\eta^+\p\xi^--\eta^-\p\xi^+]\\
&-2\eta^+e^{\phi^+}\eta^-e^{\phi^-}b-\p(\eta^-e^{\phi^-})\eta^+e^{\phi^+}\tilde b+\p(\eta^+e^{\phi^+})\eta^-e^{\phi^-}\tilde b\\
&{3\over 4}\p c(\p\phi^++\p\phi^-)+\tilde c (\p\phi^--\p\phi^+)
\ea
\eeq
Below are  expressions for BRST variations
of certain operators in the ghost CFT
\beq
\bega{ll}
\{Q,c\}=c\p c-2\gamma^-\gamma^+\\
\{Q,\gamma^\pm\}=c\p\gamma^\pm-{1\over 2}\gamma^\pm\pm\tilde c\gamma^\pm\\
\{Q,\tilde c\}=c\p\tilde c+2\gamma^-\p\gamma^+-2\gamma^+\p\gamma^-\\
\{Q,\tilde b c\}=2\gamma ^-\gamma^+\tilde b+cJ-c(\p\phi^+-\p\phi^-)\\
\{Q,e^{-\phi^+}e^{-\phi^-}\}=\p ce^{-\phi^+}e^{-\phi^-}+c\p(e^{-\phi^+}e^{-\phi^-})\\
\{Q,e^{\phi^+}e^{-\phi^-}\}=-\p ce^{\phi^+}e^{-\phi^-}+c\p(e^{\phi^+}e^{-\phi^-})+\eta_+e^{2\phi^+}e^{-\phi^-}G^-+2\tilde c e^{\phi^+}e^{\phi^-}
+2\eta^-\eta^+e^{2\phi^+}\tilde b\\
\ea
\eeq
%

\section{Superparafermions and their fusion rules}
\setcounter{equation}{0}
Here we summarize some basic facts on the $SL(2)/U(1)$ and $SU(2)/U(1)$ superparaferm\-ions
introduced in section 2.
(The two theories are closely related, and most of the formulae
differ by the substitution $k\ra -k$).
In the following we will suppress antiholomorphic terms.
The R-current of $\NN=2$ algebra of superparafermions can be bosonised as
\beq
   J_R=i\sqrt{c\over 3}\p X_R \, .
\eeq
The expressions for $\NN=2$ primaries in the R sector are \cite{Aharony:2003vk}
\beq\label{dimslrr}
  \Delta[V^{(sl,susy)}_{h;m}(R,\pm)]=-{h(h-1)\over k} + {(m\pm{1\over 2})^2\over k}+{1\over 8}
\eeq
and
\beq
\label{uonerslrr}
  q_{sl,R}=\pm {1\over2}+{2 m\pm1\over k}
\eeq
The corresponding $SU(2)/U(1)$ formulae are
\beq\label{dimsurr}
  \Delta[V^{(su,susy)}_{h;m}(R,\pm)]={j(j+1)\over k} - {(m\pm{1\over 2})^2\over k}+{1\over 8}
\eeq
and
\beq
\label{uonersurr}
  q_{su,R}=\pm {1\over2}-{2 m\pm1\over k}
\eeq

The expressions for the conformal dimensions and R-charges of parafermion
vertex operators in the NS sector are given by (\ref{dimsu})--(\ref{uonersl}). 
The superconformal primaries ${\cal V}_{h,m}'$ are related to the primaries of the
bosonic $SL(2)/U(1)$ theory (see e.g. \cite{Aharony:2003vk})
\beq
 {\cal V}_{h,m}'=V^{(sl)}_{h;m} e^{i \alpha_m X_R}, \qquad \alpha_m={2 m\over\sqrt{k(k+2)}}
\eeq
Primaries of the bosonic $SL(2)/U(1)$ theory are related to the primaries
of $SL(2)_{k-2}$ WZW model $\Phi_h(x,{\bar x})$ through the coupling to the extra $U(1)$ whose bosonised
current we denote by $\p Y$.
In addition, there is an analog of Fourier transform, which diagonalizes
$\Phi_h(x,{\bar x})$ in the $J^{sl}_3$ basis (eigenvalue of  $J^{sl}_3$ is conventionally
denoted by $m$).
\beq
  V^{(sl)}_{h;m,\bar m}=e^{i\sqrt{2\over k}(m Y-{\bar m}{\bar Y})}
                        \int d^2x x^{h+m-1} {\bar x}^{h+\bar m-1} \Phi_h(x,{\bar x})
\eeq
The construction of $SU(2)/U(1)$ parafermions is very similar and
can be found for example in \cite{Zamolodchikov:1986gh,Qiu:1986zf}

The $SU(2)$ fusion rules  are
\begin{equation}\label{opek3}
[\CV_{\frac{k-2}{2},m}][\CV_{\frac{k-2}{2},m'}]\sim [\CV_{0,m+m'}] =
\left\{ \begin{array}{ll}[G^{-}_{-1/2}\CV_{\frac{k-2}{2},m + m' - k/2}]&
\mbox{if} \enspace k-2> m+m' > 0\\
 {[} G ^{+}_{-1/2}\CV_{\frac{k-2}{2},m + m' + k/2}] & \mbox{if} \enspace 2-k< m+m' < 0 \\
 1& \mbox{if}\enspace m + m'=0
\end{array} \right.
\end{equation}
while in  $SL(2)$ we have \cite{Ahn:2004qb,Hosomichi:2004ph}
\begin{equation}\label{opek4}
\bega{ll}
[\CV'_{-k/2, m}][\CV'_{-k/2, m'}] \sim [G^{-}_{-1/2}\CV'_{-k/2, m+m'+k/2}] + [\CV'_{-k,m+m'}]
+ [\CV_{0,m+m'}]  \enspace \\
\\
\qquad\qquad\qquad\qquad\mbox{if}\enspace 0> m+m'>-(k-2) \, .
\ea
\end{equation}
 For integral level $k$, that
is the case of interest for the minimal $N=2$ strings, a special degeneration occurs
in the $SL(2)/U(1)$ OPE \ref{opek4} - the family $[\CV'_{-k,m+m'}]$ drops out of the theory
(see e.g. \cite{Teschner}).

In general if we have an OPE of the type
\beq
\CVp_{1}(z)\CVp_2(0)\sim C G^-_{-1/2} \CVp_3(0).
\eeq
with the fusion coefficient $C$, it turns out that the fusion coefficients in the following OPE's
\beq
\bega{ll}
G^+_{-1/2}\CVp_{1}(z)\CVp_2(0)\sim C_1\CVp_3(0)\\
\CVp_{1}(z)G^-_{-1/2}\CVp_2(0)\sim C_2\CVp_3(0),
\ea
\eeq
are
\beq
\label{strube}
C_1=-C_2=-C(2\Delta_{3}+q_{3}).
\eeq

We will also need to know the fusion coefficients in the following OPE
\beq
G^-_{1/2}\CVp_1(z)\CVp_2(0)\sim C_3 G^-_{-3/2}G^-_{-1/2}\CVp_3(0)
\eeq
To compute it we consider the following $3$-pt function
\footnote{Note that $G^{-}_{-3/2}G^{-}_{-1/2}\CVp_3$ is a Virasoro primary and hence has simple scaling behavior at infinity.}
\beq
\bega{ll}
\langle G^-_{-1/2}\CVp_1\CVp_2 G^+_{-3/2}G^+_{-1/2}\CVp_3\rangle=
\langle G^-_{-1/2}\CVp_1 G^+_{-1/2}\CVp_2 G^+_{-1/2}\CVp_3\rangle\left({1\over z_1-z_3}-{1\over z_2-z_3}\right)-\\
\\
{(2\Delta_1+q_1)\over(z_1-z_3)^2}\langle \CVp_1\CVp_2 G^+_{-1/2}\CVp_3\rangle
\ea
\eeq
This expression allows us to find also the following fusion coefficient
\beq
G^-_{-1/2}\CVp_1 G^+_{-1/2}\CVp_2\sim C_4 G^-_{-1/2}\CVp_3
\eeq
We find
\beq\label{structsl}
\bega{ll}
C_4+C (2\Delta_1+q_1)=0\\
C_3(2\Delta_3+3q_3-2+2\tilde c)=-C(2\Delta_1+q_1)
\ea
\eeq

The expressions above are derived for the fusion rules (\ref{opek4})
which are relevant for the $SL(2)/U(1)$ parafermions.
Similar expressions hold for their $SU(2)/U(1)$ counterparts;
one is instructed to invert the sign of the $U(1)_R$ charge.

There is an alternative way to derive the fusion rules and the
structure constants discussed above.
It involves considering the null vectors and their effect on the OPEs.
Let's start with the simplest case of level one null vector.
As discussed above the null vector has the following form
\beq
[(q_1-1)L_{-1}-(2\Delta_1+1)J_{-1}+G^{+}_{-1/2}G^{-}_{-1/2}]|\Delta_1,q_1\rangle\equiv \hat f|\Delta_1,q_1\rangle,
\eeq
where $\Delta_1$ and $q_1$ satisfy
\beq
2(\tilde c -1)\Delta_1=q_1^2-\tilde c.
\eeq
We will be interested in OPE's of the following type (we suppress the $z$ dependence)
\beq\label{opegp}
\bega{ll}
\CVp_1(z)\CVp_2(0)\sim CG^{+}_{-1/2}\CVp_3(0)\\
\\
G^-_{-1/2}\CVp_1(z)G^+_{-1/2}\CVp_2(0)\sim C_4 G^{-}_{-1/2}\CVp_3(0)
\ea
\eeq
To find the allowed $\CVp_3$ we consider the following correlator
\beq\label{nullope}
\langle \hat f\CVp_1(z_1)\CVp_2(z_2)G^{+}_{-1/2}\tilde\CVp_3\rangle=0,
\eeq
where $\tilde\CVp_3$ is the conjugate of $\CVp_3$.
Expanding (\ref{nullope}) we find
\beq\bega{ll}
\left[(q_1-1)\p_{z_1}-(2\Delta_1+1)\left(-{q_2\over z_{21}}+{q_1+q_2\over z_{31}}\right)\right]
\langle\CVp_1(z_1)\CVp_2(z_2)G^{+}_{-1/2}\tilde\CVp_3\rangle=\\
\\
\qquad\qquad-\langle G^-_{-1/2}\CVp_1(z_1)G^+_{-1/2}\CVp_2(z_2)G^{+}_{-1/2}\tilde\CVp_3\rangle,
\ea
\eeq
where $z_{ij}=z_i-z_j$.
Using this equation we find the following relations
\beq
\bega{ll}
(q_1-1)\left(\Delta_{21}-{1\over 2}\right)C=-(2\Delta_1+1)q_2 C+C_4\\
\\
\left(\Delta_{31}+{1\over 2}\right)(q_1-1)=(2\Delta_1+1)(q_1+q_2),
\ea
\eeq
where $\Delta_{21}=\Delta_2+\Delta_1-\Delta_3$ and $\Delta_{31}=\Delta_3+\Delta_1-\Delta_2$.
One can check that these equations are in complete agreement with (\ref{opek4}) and (\ref{structsl}).
In terms of $h,m$ quantum numbers (\ref{opegp}) can be written as follows
\beq\label{opegpex}
\CVp_{-{k\over 2},m_1}\CVp_{h,m_2}\sim CG^-_{-1/2}\CVp_{h,m_1+m_2+k/2}.
\eeq

The coefficient $C$ in (\ref{opegp}) will become zero for $k$ and $k-m_1$ integer, since two additional null vectors associated with the
zeros of polynomial (\ref{gseries}) will appear. This problem can be dealt with by starting from irrational $k$
and redefining the $SL(2)/U(1)$ operators in such a way to make $C$ finite.

\section{Fusion of the $\Or^{(1)}_{m}$ operators.}
\setcounter{equation}{0}
Here we give details of the fusion of operators $\Or^{(1)}_m$ constructed in section 4.
We start by fixing the normalization of the level one SL(2)/U(1) operators.
Consider the OPE involving the operator $\CVp^{(1)}_m$, with $m\in \IR$  and a generic operator
$\CVp_{h,m'}$,
\begin{eqnarray}\label{fusionone}
\CVp^{(1)}_m(z)\CVp_{h,m'}(0)\sim C_{\uparrow}\left(h-m-m'+{k\over 2}\right)^{-1}G^+_{-1/2}\CVp_{h,m+m'-{k\over 2}}+\nonumber\\
C_{\downarrow}\left(h+m+m'+{k\over 2}\right)^{-1}G^-_{-1/2}\CVp_{h,m+m'+{k\over 2}}+\cdots,
\end{eqnarray}
where we suppressed the $z$-dependence and the ellipses stand for two additional terms in the OPE, which are not relevant
for our consideration.
The structure constants $C_{\uparrow}$ and $C_{\downarrow}$ can be found in e.g. in \cite{Hosomichi:2004ph}.
Up to an uninteresting overall multiplicative factors these are
\begin{eqnarray}
C_{\uparrow}^2={\Gamma(1-h-m')\Gamma(1+{k\over 2}-m)\Gamma(h+m+m'-{k\over 2}-1)
\over \Gamma(h+m')\Gamma(-{k\over 2}+m)\Gamma(2-h-m-m'+{k\over 2})} \\
C_{\downarrow}^2={\Gamma(1-h+m')\Gamma(1+{k\over 2}+m)\Gamma(h-m-m'-{k\over 2}-1)
\over \Gamma(h-m')\Gamma(-{k\over 2}-m)\Gamma(2-h+m+m'+{k\over 2})}
\end{eqnarray}
We immediately see that both $C_{\uparrow}$ and $C_{\downarrow}$  vanish for
\beq\label{mone}
m=-{k\over 2},\,-{k\over 2}+1,\,\cdots,\, {k\over 2},
\eeq
as expected.
Redefining the SL(2)/U(1) operators as 
\beq
\tilde\CVp_m\equiv\lim_{\epsilon\rightarrow 0} \CVp_{m+\epsilon}\sqrt{\Gamma\left (-{k\over 2}+m+\epsilon\right)},
\eeq
we see that both $\tilde C_{\uparrow}$ and $\tilde C_{\downarrow}$ become finite for
$m$ satisfying (\ref{mone}).
Note that the case of $h=-k/2$ and $m+m'=0$ is special since the coefficients in (\ref{fusionone}) blow up.
At this point we need to switch our attention to the whole operator $\Or^{(1)}_m$.
There is an additional factor $K_{m_1,m_2}$ in the fusion of  $\Or^{(1)}_m$'s
\beq
PCO^{+}\Or^{(1)}_{m_{1}}PCO^{-}\Or^{(1)}_{m_{2}} = C_{m_{1},m_{2}}K_{m_{1},m_{2}}\Or^{(1)}_{m_{3}} \, .
\eeq
Here $m_{3} = m_{1} + m_{2} -k/2$ and $C_{m_{1},m_{2}} \propto C_{\uparrow}(m_{1},m_{2})\left(h-m_{1}-m_{2}+{k\over 2}\right)^{-1} $ for
$m_{1} + m_{2}>0$ and  $m_{3} = m_{1} + m_{2} +k/2$, $C_{m_{1},m_{2}} \propto C_{\downarrow}(m_{1},m_{2})\left(h+m_{1}+m_{2}+{k\over 2}\right)^{-1} $ for
$m_{1} + m_{2}<0$. For $m_{1}+m_{2}=0$ the corresponding fusion rule is defined as follows. First of all note that the relevant $SU(2)$ fusion
rule has the form
\beq
[{\cal V}_{m}^{(1)}][{\cal V}_{-m}^{(1)}] \sim [1]
\eeq
so that both terms in the right hand side of (\ref{fusionone}) contribute to
\beq\label{id}
{\cal O}^{(1)}_{-\frac{k}{2}}={\cal O}^{(1)}_{\frac{k}{2}}=c\partial\xi^{+} \CVp_{-\frac{k}{2}}^{(1)}e^{-2\phi_{-}}e^{-\phi_{+}}  +
\partial\xi^{-} c\CVp_{\frac{k}{2}}^{(1)} e^{-2\phi_{+}}e^{-\phi_{-}}\, .
\eeq
The relevant proportionality coefficient $K_{m,-m}$ in this case vanishes precisely in such a way that the product
$C_{m_{1}, m_{2}}K_{m_{1},m_{2}}$ becomes finite. The operator (\ref{id}) acts as an identity operator within the ground ring.


\section{ Action of the picture raising operators on BRST  cohomologies}
\setcounter{equation}{0}
Here we give an example of a kernel and a cokernel of a picture-raising operator
(\ref{pco}) considered on BRST cohomologies.
We begin with an example of a situation when   $PCO^\pm$  map  BRST non-trivial operators into
BRST trivial ones. The simplest  example of such phenomenon is the operator
\beq
{\cal O}=ce^{-\phi_-}e^{-\phi_+}{\bf 1}~.
\eeq
This operator is mapped into zero by the action of both $PCO^\pm$. In general BRST non-trivial
operators of the type
\beq
{\cal O}=ce^{-\phi_-}e^{-\phi_+}{\cal V}{\cal V}^\prime~,
\eeq
where $\cal V$ and ${\cal V}^\prime$ are (anti-)chiral primaries
of $SU(2)/U(1)$ and $SL(2)/U(1)$ respectively
are mapped into zero by the action of $PCO^+$
($PCO^-$)\footnote{It should be noted, however, that in contrast with
the case considered in section 2, here the image of the $PCO^\pm$
is not an exact zero but rather a null state.}


Consider now a  BRST invariant operator in the standard $(-1,-1)$ picture
\beq
{\cal O}=c e^{-\phi_-}e^{-\phi_+}{\cal V}^\prime {\cal V}\, .
\eeq
The operator ${\cal V}$ and ${\cal V}^\prime$ should be primaries
of the corresponding ${\cal N}=2$
algebras, in order for $\cal O$ to be BRST closed.
Now let us perform picture changing into the $(0,-1)$ picture. The result is
\beq\label{pr}
{\cal O}^\prime=ce^{-\phi_-}G^{-}_{-{1\over 2}}({\cal V}^\prime {\cal V})\, .
\eeq
In general the BRST cohomologies in this picture have the following form
\beq \label{pr2}
\tilde{\cal O}=ce^{-\phi_-}{\tilde{\cal V}}^\prime {\tilde{\cal V}}~,
\eeq
where ${\tilde{\cal V}}^\prime {\tilde{\cal V}}$ satisfies
\beq
\bega{ll}
G^{+}_r({\tilde{\cal V}}^\prime {\tilde{\cal V}})=0~,\\
G^{-}_{r-1}({\tilde{\cal V}}^\prime {\tilde{\cal V}})=0\quad r>0~,\\
\Delta({\tilde{\cal V}}^\prime {\tilde{\cal V}})={1\over 2}~,\\
q(({\tilde{\cal V}}^\prime {\tilde{\cal V}})=-1~.
\ea
\eeq
These conditions are indeed satisfied for the ${\cal O}^\prime$, but it is easy to see that
in this picture there are additional BRST invariant operators corresponding to the fields
which are antichiral in $SU(2)/U(1)$ and $SL(2)/U(1)$ separately. This operator is based on primaries
in $SU(2)/U(1)$ and $SL(2)/U(1)$ which
are labeled the following quantum numbers
\beq
\bega{ll}
j^{su}=m^{su}={k\over 2}-h~,\\
m^{sl}=-h~,\\
h^{sl}=h~.
\ea
\eeq
A simple combinatorial analysis, similar to the one done in  section \ref{gsec} of the paper,
 shows that one cannot write down an operator
with ghost number one, picture $(-1,-1)$, vanishing total $R$-charge and dimension,
which would be based on a matter primary with dimension $1/2$ and $R$-charge -1.
Thus there is no  $PCO^{+}$ pre-image for operator (\ref{pr2}).

\end{document}